\documentclass[a4paper]{aa}
\usepackage{graphicx}
\usepackage{txfonts}
\begin{document}
\title{Non-axisymmetric instability of axisymmetric magnetic fields} 
\author{A.~Bonanno\inst{1,2}, V.~Urpin\inst{1,3}}
%\offprints{}
\institute{$^{1)}$ INAF, Osservatorio Astrofisico di Catania,
           Via S.Sofia 78, 95123 Catania, Italy \\
           $^{2)}$ INFN, Sezione di Catania, Via S.Sofia 72,
           95123 Catania, Italy \\
           $^{3)}$ A.F.Ioffe Institute of Physics and Technology and
           Isaac Newton Institute of Chile, Branch in St. Petersburg,
           194021 St. Petersburg, Russia}
\date{\today}

\abstract
{The MHD instabilities can generate complex field topologies even if the 
initial field configuration is a very simple one.}
{We consider the stability properties of magnetic configurations containing 
a toroidal and an axial field. In this paper, we concentrate mainly on the
behavior of non-axisymmetric perturbations in axisymmetric magnetic 
configurations.} 
{The stability is treated by  a linear analysis of ideal MHD
equations.} 
{In the presence of an axial field, it is shown that the instability can
occur for a wide range of the azimuthal wavenumber $m$, and its growth rate
increases with increasing $m$. At given $m$, the growth rate  is at its maximum for
perturbations with the axial wave-vector that makes the Alfv\'en frequency
approximately vanishing. We argue that the instability of magnetic 
configurations in the ideal MHD can typically be dominated by perturbations 
with very short azimuthal and axial wavelengths.}
{}


\maketitle

\section{Introduction}  
A wide variety of MHD instabilities can occur in magnetized astrophysical
bodies where they play an important role in the evolution and formation of 
various structures and in enhancing transport processes, among others. The onset of 
instabilities can be caused both by hydrodynamic motions (for instance, 
differential rotation) or properties of the magnetic configuration. Even
magnetic fields with a relatively simple topology (for example, a purely 
toroidal field) can be subject to instability. Magnetic fields generated by 
the dynamo action or stretched by hydrodynamic motions are topologically 
more complex and can cause this sort of instability as well. Which field 
strength and topology can sustain a stable magnetic configuration is still 
rather uncertain despite all the extensive work already done (see Borra et al. 1982; Mestel 
1999 for review). 

The simplest and best-studied magnetic configuration is most likely a purely 
toroidal one. {This has been known since the paper by Tayler (1957), where stability
properties of the toroidal field $B_{\varphi}$ are determined by the parameter 
$\alpha = d \ln B_{\varphi} / d \ln s$  where $s$ is the cylindrical radius. 
The field is unstable to axisymmetric perturbations if $\alpha  > 1$ and to 
non-axisymmetric perturbations if $\alpha> -1/2$.} The growth time of 
instability is close to  the time taken for an Alfv\'en wave to travel 
around the star on a toroidal field line. Numerical modeling by Braithwaite 
(2006) confirms that the toroidal field with $B_{\varphi} \propto s$ or 
$\propto s^2$ is unstable to the $m=1$ mode ($m$ is the azimuthal wave number)
as  predicted by Tayler (1957, 1973). However, even a purely toroidal 
field can be stable in the region where it decreases rapidly with $s$. 
{A purely toroidal field cannot be stable through the whole star 
because the stability condition for axisymmetric modes ($\alpha < 1$) is 
incompatible with the condition that the electric current in the $z$ 
direction has no singularity at $s \rightarrow 0$, which implies $\alpha > 1$.}
The stability of the toroidal field in rotating stars has 
been considered by Kitchatinov \& R\"udiger (2007), who argue that the 
magnetic instability is essentially three-dimensional and that the finite 
thermal conductivity creates a strong destabilizing effect. Terquem \& 
Papaloizou (1996) and Papaloizou \& Terquem (1997) considered the stability 
of an accretion disk with the toroidal magnetic field and found 
that the disks containing a purely toroidal field are always unstable and 
calculated the spectra of unstable modes in the local approximation.

The stability properties of purely poloidal magnetic fields are 
also well-studied. It has been understood since the papers by Wright (1973) and 
Markey \& Tayler (1973, 1974) that the poloidal field is subject to 
dynamical instabilities in the neighborhood of points (or lines) { where
the poloidal field is vanishing (neutral points/lines).} 
These authors recognized first that the magnetic field in the neighborhood 
of a neutral line resembles that of a toroidal, pinched discharge, which is 
known to be unstable. The instability of a poloidal field is also rather 
fast: its growth time can reach a crossing time of few Alfv\'en times (Van Assche et 
al. 1982; Braithwaite \& Spruit 2006) that is very short, for example, 
compared to the time-scales of stellar evolution. However, the instability 
of a poloidal field can be suppressed by the addition of a toroidal field in 
the neighborhood of neutral points (Markey \& Tayler 1973; Wright 1973).

{ Conversely}, the addition of even a relatively weak poloidal field 
alters  the stability properties of the toroidal field substantially. { For 
example, if the poloidal field is uniform and relatively weak, the instability 
condition of axisymmetric modes reads $\alpha > -1$,  at variance with the
condition of instability for a purely toroidal field (see, e.g., Knobloch 1992;
Dubrulle \& Knobloch 1993), which predicts that an unstable toroidal field 
configuration has $\alpha > -1/2$. 
%In accordance with this condition, the instability
%can arise in a wider range of $\alpha$ because a purely toroidal field is
%stable for $-1/2 > \alpha > -1$. 
Therefore, a weak poloidal field has a
destabilizing effect. However, a  strong enough poloidal field can 
suppress the instability of the toroidal field.} It turns out that configurations 
containing comparable toroidal and poloidal fields are more 
stable than purely toroidal or purely poloidal ones (Prendergast 1956; 
Tayler 1980) and, generally, the possibility exists that there are 
configurations containing mixed fields, which have no instabilities arising
on a dynamical time-scale. 
In his study of 
unstable magnetic configurations Tayler (1980) has not found any instability if the axial 
field $B_z > 0.3 B_{\varphi}$ for instance, even though such configurations can be unstable for 
a wide range of the azimuthal wavenumber $m$ if $B_z$ is weaker. With 
numerical simulations Braithwaite \& Nordlund (2006) studied the stability of 
a random initial field in the stellar radiative zone and found that the stable
magnetic configurations generally have the form of tori with comparable 
poloidal and toroidal field strengths.

In this paper, we consider in detail the stability properties of magnetic 
configurations containing the toroidal and axial magnetic fields with respect 
to non-axisymmetric perturbations. We show that the instability may occur for 
a wide range of the azimuthal wavenumber $m$, and the growth rate is typically 
higher for higher $m$. Unstable modes with large $m$ have a very short vertical
lengthscale, so it can be hard to resolve them in numerical calculations. 
Depending on the profile $B_{\varphi}(s)$ and the ratio $B_z/B_{\varphi}$, the 
instability can occur in two regimes that have substantially different growth 
rates. 

The remainder of this paper is arranged as follows. In Sect. 2, we derive
the equation that governs the eigenfunctions and eigenvalues of the magnetic
field. We describe the numerical procedure and present the results of 
calculations in Sect. 3. A brief discussion of the results is given in 
Sect. 4.

\section{Basic equations}
Let us consider the stability of an axisymmetric cylindrical magnetic
configuration in a high conductivity limit. We work in cylindrical 
coordinates ($s$, $\varphi$, $z$) with the unit vectors ($\vec{e}_{s}$, 
$\vec{e}_{\varphi}$, $\vec{e}_{z}$). We assume that the azimuthal field 
depends on the cylindrical radius alone, $B_{\varphi}= B_{\varphi}(s)$, 
but the axial magnetic field $B_z$ is constant. 

In the incompressible limit, the MHD equations read  
\begin{eqnarray}
\frac{\partial \vec{v}}{\partial t} + (\vec{v} \cdot \nabla) \vec{v} = 
- \frac{\nabla p}{\rho} 
+ \frac{1}{4 \pi \rho} (\nabla \times \vec{B}) \times \vec{B}, 
\end{eqnarray}
\begin{equation}
\nabla \cdot \vec{v} = 0, 
\end{equation}
\begin{equation}
\frac{\partial \vec{B}}{\partial t} - \nabla \times (\vec{v} \times \vec{B}) 
= 0,
\end{equation}
\begin{equation}
\nabla \cdot \vec{B} = 0. 
\end{equation}
In the basic state, the gas is assumed to be in hydrostatic equilibrium, 
then
\begin{equation}
\frac{\nabla p}{\rho} = \frac{1}{4 \pi \rho} 
(\nabla \times \vec{B}) \times \vec{B}.
\end{equation}
In this paper, we consider the stability of non-axisymmetric perturbations. 
Since the basic state is stationary and axisymmetric, the dependence of 
perturbations on $t$, $\varphi$, and $z$ can be taken in the form 
$\exp{(\sigma t - i k_z z - i m \varphi)}$ where $k_z$ is the wave-vector in 
the axial direction and $m$  the azimuthal wavenumber. In stellar 
conditions, such a local analysis in the $z$-direction applies if $k_z$ 
satisfies the inequality $k_z s > 1$ (we assume that the lengthscale in the 
axial direction is $\sim s$ in stars). Small perturbations will be indicated 
by subscript 1, 
while unperturbed quantities will have no subscript. Then, the linearized 
Eqs.~(1)-(4) are
\begin{equation}
\sigma \vec{v}_{1} = - \frac{\nabla p_{1}}{\rho} + 
\frac{1}{4 \pi \rho}[ (\nabla \times \vec{B}_{1}) \times \vec{B} 
+ (\nabla \times \vec{B}) \times \vec{B}_{1}], 
\end{equation}
\begin{equation}
\nabla \cdot \vec{v}_{1} = 0, 
\end{equation}
\begin{equation}
\sigma \vec{B}_{1} - (\vec{B} \cdot \nabla) \vec{v}_{1} + (\vec{v}_{1}
\cdot \nabla) \vec{B} = 0,
\end{equation}
\begin{equation}
\nabla \cdot \vec{B}_{1} = 0. 
\end{equation}
Eliminating all variables in favor of the radial velocity perturbation 
$v_{1s}$, we obtain 
\begin{eqnarray}
\frac{d}{ds} \left[ \frac{1}{\lambda} (\sigma^2 + \omega_A^2) \left( 
\frac{d v_{1s}}{ds} + \frac{v_{1s}}{s} \right) \right] 
- k_z^2 (\sigma^2 + \omega_A^2) v_{1s}  \nonumber \\
- 2 \omega_{B} \left[ k_z^2 \omega_{B} (1 - \alpha) - \frac{m (1 + \lambda)}{s^2
\lambda^2} \left( 1 - \frac{\alpha \lambda}{1 + \lambda} \right) (\omega_{Az} +
2 m \omega_{B} ) \right.
\nonumber \\
\left. - \frac{m \omega_{Az}}{s^2 \lambda^2} \right] v_{1s} + 
\frac{4 k_z^2 \omega_{A}^2 \omega_{B}^2}{\lambda (\sigma^2 + \omega_{A}^2)} v_{1s}
=0, \;\;\;\;\;\;
\end{eqnarray} 
where
\begin{eqnarray}
\omega_{A} =  \frac{1}{\sqrt{4 \pi \rho}} \left( k_z B_z + \frac{m}{s} 
B_{\varphi} \right), \;\;
\omega_{Az} =  \frac{k_z B_z}{\sqrt{4 \pi \rho}} , \;\;
\nonumber \\
\omega_{B} \! = \! \frac{B_{\varphi}}{s \sqrt{4 \pi \rho}} , \;\;
\alpha \! = \! \frac{\partial \ln B_{\varphi}}{\partial \ln s} , \;\;
\lambda = 1 + \frac{m^2}{s^2 k_z^2} . \nonumber 
\end{eqnarray}
For axisymmetric perturbations ($m=0$), Eq.~(10) recovers Eq.~(11) of the 
paper by Bonanno \& Urpin (2007), if one assumes $B_z=$const. The equations 
derived by Acheson (1973) and Knobloch (1992) for axisymmetric perturbations 
can also be recovered from Eq.~(10). With appropriate boundary conditions, 
Eq.~(10) allows  the eigenvalues $\sigma$ to be determined. 

{ Our analysis does not include gravity, which is important in stars, so
the stabilizing effect of stratification is neglected. In the case of
magnetic instabilities, this is justified if the work done by a perturbation 
against gravity is less than the energy released from the magnetic field. 
The corresponding condition for stellar radiative zones has been considered by
Spruit (1999) (see Eq.~(44)) and reads in our notations as $k_z s > N/\omega_{B}$,
where $N$ is the buoyancy frequency. Assuming that the characteristic value of
$N$ in stars is $\sim 10^{-3}$ s$^{-1}$ and introducing the axial wavelength
$\lambda_z =2 \pi/ k_z$, we obtain that the effect of stratification can be 
neglected for perturbations with
\begin{equation}
\lambda_z < \lambda_c = 2 \times 10^8 B_{\varphi 3} \rho_{-4}^{-1/2} \;\; 
{\rm cm},
\end{equation}
where $B_{\varphi 3}= B_{\varphi}/ 10^3$ G and $\rho_{-4} = \rho/ 10^{-4}$
g/cm$^3$.

Magnetic perturbations can also be affected  by the presence of dissipation, which is
not considered in our analysis. Usually, magnetic diffusion is weak in 
stellar conditions, and its influence is unimportant if the dissipation rate 
that is $\sim \eta k_z^2$ is small compared to $\omega_B$ when $\eta \sim 10^3$ 
cm$^2$/s  the magnetic diffusivity. This condition yields
\begin{equation}
\lambda_z > \lambda_d = 3 \times 10^5 B_{\varphi 3}^{-1/2} s_{11}^{1/2} 
\rho_{-4}^{-1/4} \;\;\; {\rm cm},
\end{equation} 
where $s_{11} = s/10^{11}$ cm. Inequalities (11) and (12) set a range of 
axial wavelengths where our consideration is valid. 
}

\section{Numerical results}
We assume  that the dependence of the azimuthal magnetic field on $s$ is 
given by
\begin{equation}
B_{\varphi} = B_{\varphi 0} \left( \frac{s}{s_1} \right)^{\alpha},
\end{equation} 
where $B_{\varphi 0}$ is the field strength at $s=s_1$. To calculate the 
growth rate of the instability, it is  convenient to introduce dimensionless 
quantities
\begin{equation}
x \!= \! \frac{s}{s_1},\;\; q \!= \! k_z s_1,\;\; \Gamma \! = \!
\frac{\sigma}{\omega_{B0}}, \;\; \omega_{B0} \!= \! \frac{B_{\varphi 0}}{s_1
\sqrt{4 \pi \rho}}, \;\;
\varepsilon \! = \! \frac{B_z}{B_{\varphi 0}}.  
\end{equation}
Then, Eq.~(10) transforms into
\begin{eqnarray} \label{pert}
\frac{d}{dx} \left( \frac{d v_{1s}}{dx} \! + \! \frac{v_{1s}}{x} \right) +
\left( \frac{d v_{1s}}{dx} \! + \! \frac{v_{1s}}{x} \right) \frac{d \ln \Delta}{d x}
\! - \! q^2 \left(1 + \frac{m^2}{q^2 x^2} \right) v_{1s} -
\nonumber \\
\frac{2 q^2 x^{\alpha -1}}{\Gamma^2 + f^2} \left\{ \left[ \left( 1 \! - \!
\frac{m^2}{q^2 x^2} \right) x^{\alpha -1} \! - \! \frac{m \varepsilon}{q x^2} 
\right]
(1 - \alpha) \! - \! \frac{2 m f}{m^2 + q^2 x^2} \right\} v_{1s} 
\nonumber \\
+ \frac{4 q^2 f^2 x^{2(\alpha -1)}}{(\Gamma^2 + f^2)^2} v_{1s} = 0, \;\;\;\;\;\;
\end{eqnarray}
where
\begin{equation}
f = q \varepsilon + m x^{\alpha -1}, \;\; \Delta = \frac{q^2 x^2 (\Gamma^2 +
f^2)}{m^2 + q^2 x^2}.  
\end{equation}
Solving  the eigenvalue problem numerically, we assume that the inner and outer
boundaries correspond to $x_1 =1$ and $x_2 =2$, respectively, where we suppose 
$v_{1s}(x_1)=v_{1s}(x_2) =0$. { Note, however, that the results are not 
qualitatively sensitive to the choice of boundaries and boundary conditions. 
We checked that the qualitative behavior of the growth rate does not change
if the outer boundary $x_2$ is set to higher values. As far as the inner 
boundary is concerned, our approach does not allow  the 
eigenvalues to be calculated in the case $s_1 = 0$, but we verified that low but finite
values of $s_1$ did not lead to qualitative difference in the behavior 
of the growth rate, and therefore we  believe that our  
conclusions  can be extended to the case $s_1 = 0$.}

Equation~(15),  together with the given boundary conditions at the extrema, is a 
two-point boundary value problem that can be solved by using the ``shooting''
method (Press et al., 1992). To solve Eq.~(15), we used a 
fifth-order Runge-Kutta integrator embedded in a globally convergent 
Newton-Rawson iterator. We have checked that the eigenvalue was always the 
fundamental one, as the corresponding eigenfunction had no zero 
except that at the boundaries. 
\begin{figure}
%\begin{center}
\includegraphics[width=9.0cm]{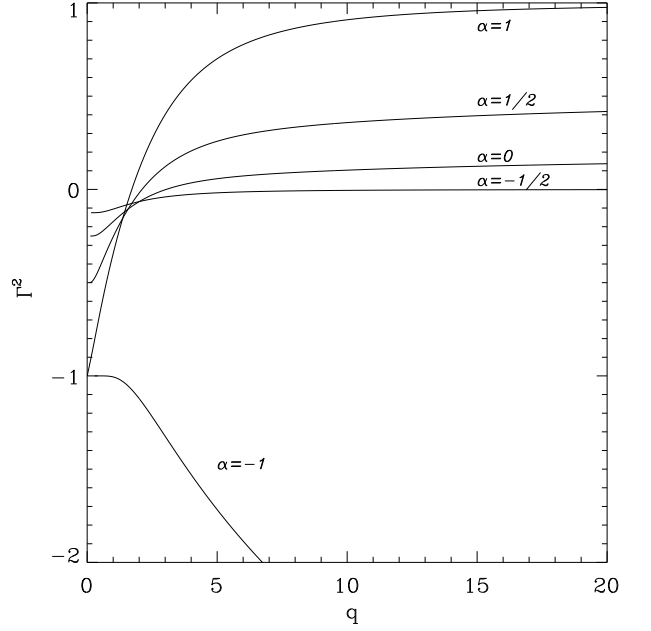}
\caption{The dependence of $\Gamma^2$ on $q$ for $B_z=0$ and $m=1$.
Numbers near the curves indicate the values of parameter $\alpha$.}
%\end{center}
\end{figure}
To test the numerical procedure, we consider first the case of a purely
toroidal field, $\varepsilon = 0$. Stability properties of such field have been
well studied since the paper by Tayler (1973), who argued that the dominant
unstable mode is the one  with $m=1$. The instability of the toroidal field occurs
if $d (s B_{\varphi}^2) / ds > 0$ or, in other words, if $B_{\varphi}$ 
increases with the cylindrical radius or decreases not faster than 
$s^{-1/2}$. In Fig.~1, we plot the growth rate as a function of the vertical 
wave-vector for $B_z=0$ and different values of $\alpha$. In agreement with 
the conclusion by Tayler (1973), $\alpha = - 1/2$ is indeed the critical 
value distinguishing between stable ($\alpha < - 1/2$) and unstable ($\alpha >
- 1/2$) toroidal magnetic fields. The growth rate of instability can be 
rather high, $\sigma \sim \omega_{B0}$. Note that the growth rate increases 
with increasing $q$, and perturbations with short axial wavelengths turn
out to grow the most rapidly.  
  
\begin{figure}
%\begin{center}
\includegraphics[width=9.5cm]{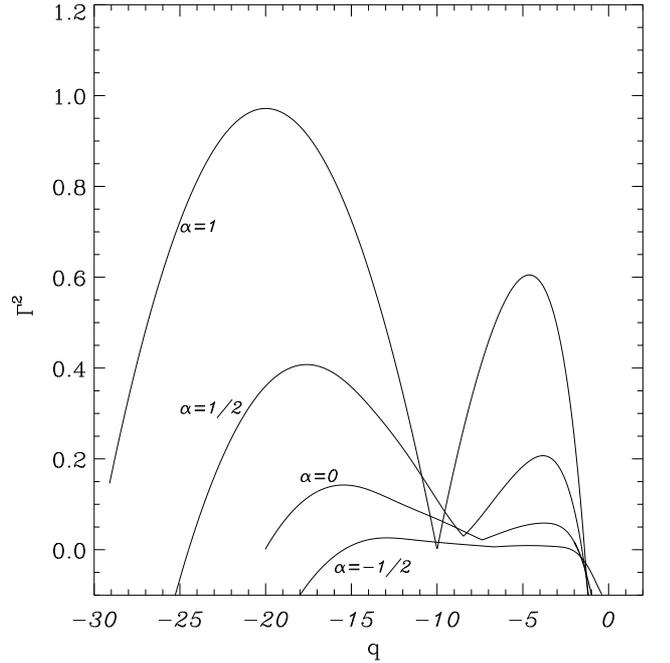}
\caption{The dependence of $\Gamma^2$ on $q$ for $\varepsilon = B_z/
B_{\varphi 0}= 0.1$ and $m=1$. Numbers near the curves correspond to
values of $\alpha$.}
%\end{center}
\end{figure}

In Fig.~2, we plot the dependence of $\Gamma^2$ on $q$ for the same $m=1$ 
mode but in the presence of a relatively weak axial field, $B_z = 0.1 
B_{\varphi 0}$. The addition of even very weak $B_z$ changes  
the stability properties qualitatively even though the energy contained in the axial field is 
very low compared to that of the toroidal field ($\sim 1$\%). 
{ We stress that the presence of an axial field breaks symmetry 
$q\rightarrow -q$. 
%positive and negative values of $q$ because the solution of Eq.~(15) depends 
%on the sign of $q$ and $m$ but not only on their values. 
The physical reason for this is fairly simple. The Lorentz force plays a crucial role in the behavior of 
perturbations, and this force contains a component that is proportional to
the cross production of a current flowing in the basic state $\vec{j} \propto 
\nabla \times \vec{B}$ and the magnetic field of perturbations (see the last 
term on the left hand side of Eq.~(6)). If $B_z \neq 0$, then magnetic perturbations 
are determined partly by the axial gradient of velocity perturbations since 
$d \vec{B}_1/dt = (\vec{B} \nabla) \vec{v}_1$, and this contribution has a
different sign for positive and negative $k_z$. Therefore, the stability
properties turn out to be dependent on the direction of an axial wavevector.
However, Eq.~(15) still contains some degeneracy because the replacements
$(m, q) \rightarrow (-m, -q)$ or $(m, \varepsilon) \rightarrow (-m, 
-\varepsilon)$ do not change its shape. }
The instability occurs only for a restricted range of negative $q$, $-(20-30) 
< q < -2$ depending on the value of $\alpha$, and do not appear for 
positive $q$. The growth rate has two clear maxima with the higher maximum 
corresponding to $q \sim - 1/ \varepsilon$. By the order of magnitude, the 
axial wave-vector of the most rapidly growing perturbation can be estimated 
from the condition 
\begin{equation}
\omega_A = \frac{1}{\sqrt{4 \pi \rho}} \left( k_z B_z + \frac{m}{s} 
B_{\varphi} \right) \approx 0.
\end{equation}     
Indeed, this equation in dimensionless units reads $\omega_A \propto q 
\varepsilon + m x^{\alpha -1} \approx 0$. Since $x \sim 1$ in our 
calculations, the condition $\omega_A = 0$ corresponds to 
\begin{equation}
q \sim - m /\varepsilon.
\end{equation} 
Therefore, we have $q \sim -1/\varepsilon$ for the $m=1$ mode. The most 
rapidly growing modes turns out to be highly anisotropic if the axial field 
is weak compared to the toroidal one.  Their axial wavelength $\lambda_z = 2 
\pi/k_z \sim 2 \pi \varepsilon s$ is much shorter than the radial and 
azimuthal lengthscale. The presence of an axial field shifts the threshold 
of unstable values of $\alpha$: a purely toroidal field is unstable only if 
$\alpha > -1/2$, but our calculations show that the profile of $B_{\varphi}$ 
with $\alpha= -1/2$ is unstable. However, the growth rate in the case $\alpha 
\leq -1/2$ is rather low and is close to $\omega_{Az}$ rather than 
$\omega_{B0}$. { The growth rate grows rapidly with increasing $\alpha$:
the maximum value of $\Gamma^2$ is of the order of unity for $\alpha =1$ and
$\sim 8$ for $\alpha=2$ (is not shown in Fig.~2).
} 

\begin{figure}
%\begin{center}
\includegraphics[width=9.5cm]{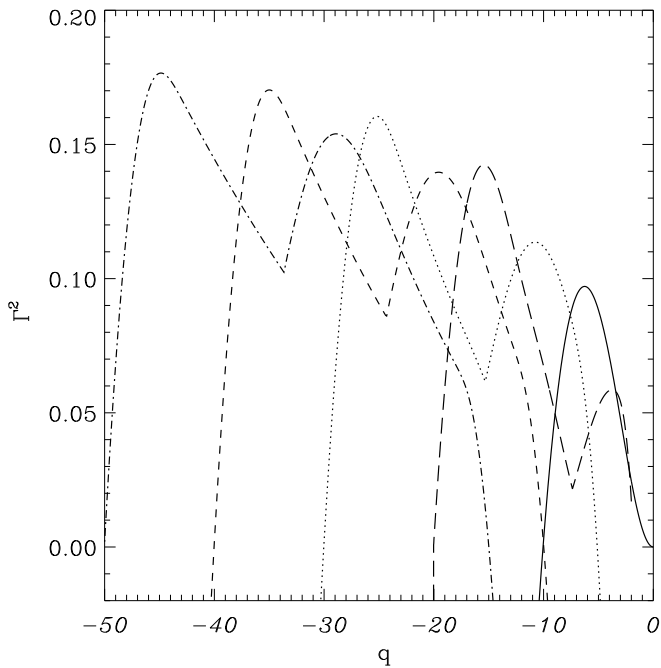}
\caption{The dependence of $\Gamma^2$ on $q$ for $\varepsilon = B_z/
B_{\varphi 0}=0.1$, $\alpha = 0$, and different values of $m$. 
Solid, long-dashed, dotted, short-dashed, and dot-dashed lines are for 
$m=0,1,2,3,4$, respectively}
%\end{center}
\end{figure}

The dependence of $\Gamma^2$ on $q$ in the presence of a weak axial field 
($\varepsilon = 0.1$) and for different values of $m$ is shown in Fig.~3. 
In this figure, $\Gamma^2$ is calculated for $\alpha = 0$, such that the 
magnetic configuration is unstable even in the absence of the axial field. 
It turns out that this profile of $B_{\varphi}$ is unstable if $B_z \neq 0$, 
as well, but the properties of instability are substantially different. In 
contrast to the case of a purely toroidal field with $\alpha = 0$ where 
only the $m=1$ mode can arise (Tayler 1973), the instability occurs for modes 
with a wide range of $m$ including $m=0$. 
%%%%%%%%%%%%%%%%%%%%%%%% QUI
For  $m \neq 0$, the growth 
rate reaches its maximum approximately at $q \approx - m /\varepsilon$, which 
is equivalent to the resonant condition $\omega_A \approx 0$. The maximum 
growth rate increases with an increase 
in $m$. Therefore, the instability of such magnetic configurations is 
probably dominated  by the modes with large $m$ and extremely short axial 
wavelengths $\lambda_z \approx -2 \pi \varepsilon s/m$. 
{ This can cause difficulties in numerical simulations, because very high
resolution in the axial and azimuthal direction would be needed 
to resolve the most unstable modes. Moreover a  nonlinear interaction of these 
modes can  produce axial lengthscales that are even shorter than those 
predicted by the linear theory. }

\begin{figure}
%\begin{center}
\includegraphics[width=9.5cm]{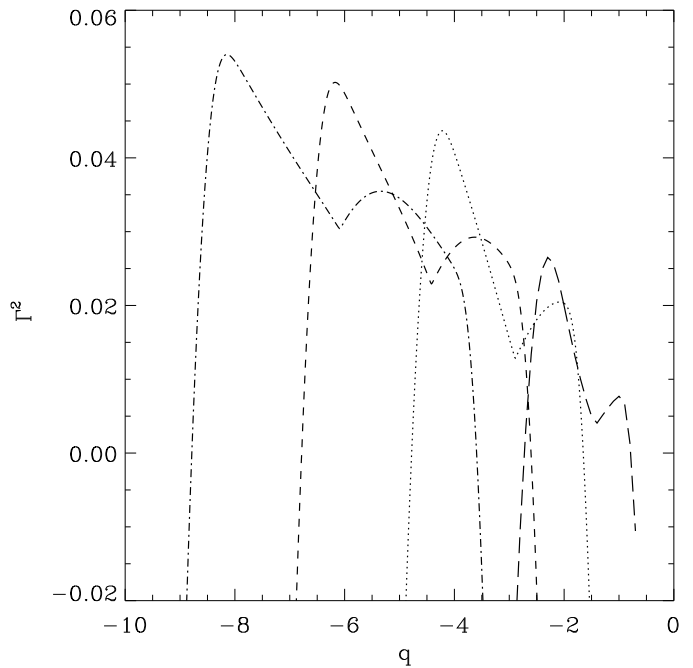}
\caption{The same as in Fig.~3 but for $B_z/B_{\varphi 0}=0.5$.}
%\end{center}
\end{figure}

An increase in the axial field makes the magnetic configuration more stable. 
In Fig.~4, we plot the growth rate for $\varepsilon=0.5$ and the same values 
of other parameter as in Fig.~3. The instability grows more slowly in higher $B_z$:
an increase in $B_z$ by a factor 5 leads to the decrease in $\Gamma^2$ 
approximately by a factor 2. Nevertheless, the instability can still occur 
for this $\varepsilon$, and it is still  efficient because its 
growth rate is $\sim 0.3 \omega_{B0}$. Like the previous case, the modes 
with a wide range of $m$ can be unstable, and the growth rate increases with 
$m$. The axisymmetric mode ($m=0$) is stable in this case (see Bonanno \&
Urpin 2007). { The  critical value $\varepsilon$ that 
suppresses the instability is clearly dependent on the geometry of the basic state. 
For example, by using energy considerations, Tayler (1980) 
has not found any instability for $\varepsilon \geq 0.3$ in the configuration 
where the magnetic surfaces of the poloidal field are coaxial tori. 
%The energy 
%consideration is explored in this analysis by asking whether any small 
%displacement reduce the potential energy of the system. 
On the other hand, for values of 
$\varepsilon$ significantly less than 0.3, the author found instability
for a wide range of values of $m$. }

\begin{figure}
%\begin{center}
\includegraphics[width=9.5cm]{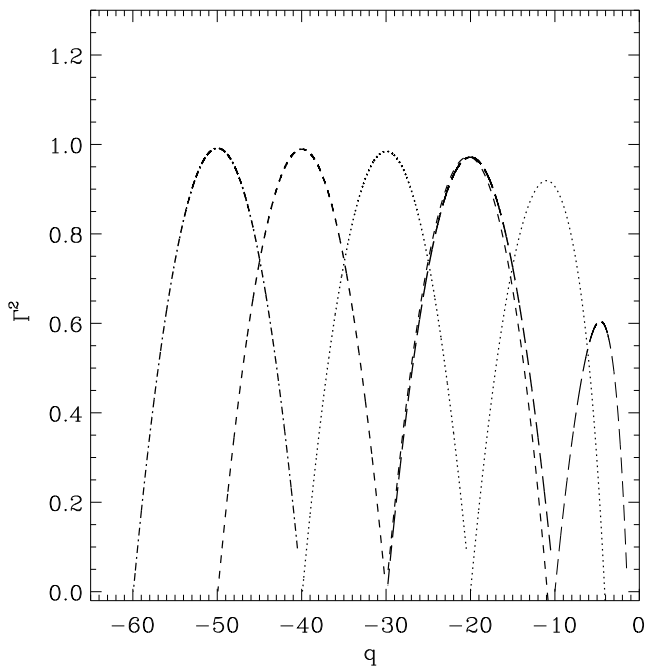}
\caption{The same as in Fig.~3 but for $\alpha = 1$.}
%\end{center}
\end{figure}

To illustrate the dependence of $\Gamma$ on the profile of the toroidal field, 
we show  the growth rate as a function of $q$ for $\alpha=1$ in Fig.~5. This
profile seems to be particularly interesting for astrophysical applications
since the toroidal field is $\propto s$ in a neighborhood of the axis of
symmetry.  In the absence of the axial field, the profile with $\alpha=1$ 
corresponds to the toroidal field that is marginally stable to axisymmetric 
perturbations ($m=0$). If $B_z \neq 0$, the threshold of instability can 
change substantially as  argued by Bonanno \& Urpin (2007) and, 
indeed, perturbations with $m=0$ turn out to be unstable in this 
case. However, the non-axisymmetric perturbations grow faster and are likely to 
dominate the instability. The maximum growth rate is very high for them and 
comparable to the Alfv\'en frequency for the toroidal field, $\omega_{B0}$. 
Remarkably,  the maximum growth rate changes rather slowly with $m$ for 
all modes with $m \geq 2$. As in the previous cases, the maximum of $\Gamma$ 
is located at $q \sim - m/ \varepsilon$. 

\begin{figure}
%\begin{center}
\includegraphics[width=9.0cm]{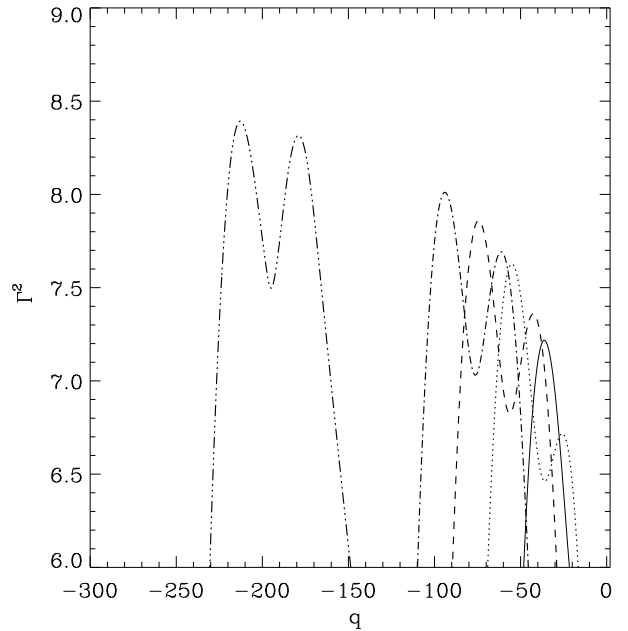}
\caption{The same as in Fig.~3 but for $\alpha = 2$. The dash-dot-dotted line
represents $m=10$.}
%\end{center}
\end{figure}

In Fig.~6, we show $\Gamma^2$ for the case of a rapid increase in 
$B_{\varphi}$ with $s$, $B_{\varphi} \propto s^2$ or $\alpha =2$. The growth 
rate of instability is substantially higher for such $\alpha$ and can reach 
the value $\sim 2-3$ toroidal Alf\'ven frequencies. A qualitative behavior of
$\Gamma$ remains same: the maximum growth rate is higher for modes with 
higher $m$, and these maxima correspond to very high values of the 
wave-vector $q \sim - m/\varepsilon$. The maximum growth rate of the
axisymmetric mode ($m=0$) is comparable to that of non-axisymmetric ones 
(see Bonanno \& Urpin 2007). Note that our calculations show some trend in 
the growth rate to reach saturation for large $m$.

\begin{figure}
%\begin{center}
\includegraphics[width=9.0cm]{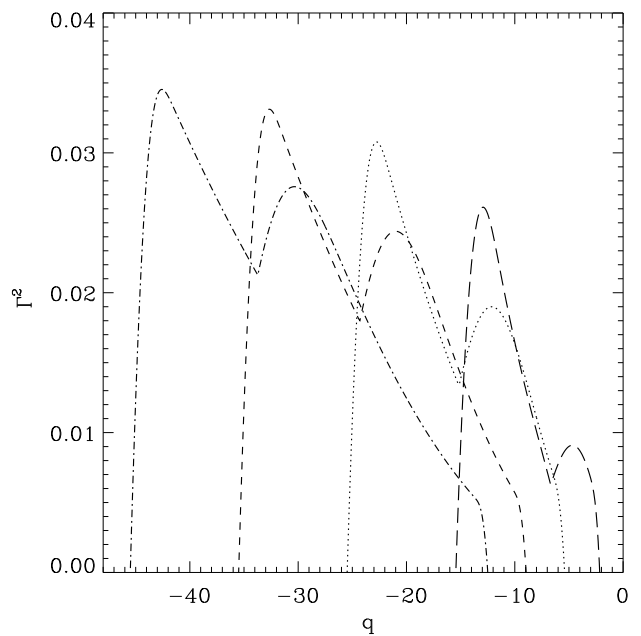}
\caption{The same as in Fig.~3 but for $\alpha =-0.5$.}
%\end{center}
\end{figure}

Figure~7 plots  the growth rate for $\alpha = - 0.5$. In accordance with Tayler 
(1973), such a profile of $B_{\varphi}$ should be marginally stable if the axial 
field is vanishing. However, the presence of $B_z$ makes  the toroidal
field unstable with such $\alpha$ even if the axial field is relatively weak. The growth
rate is not high in this case and is determined by the Alfv\'en frequency for
the axial field. As usual, the maximum growth rate is higher for higher $m$, 
and these maxima are reached for very large $q$, which corresponds to a short 
axial wavelength. A comparison between Figs.~6 and 7 illustrates the 
difference well between two regimes of the instability first noted by Bonanno \&
Urpin (2007). If $\alpha < \alpha_{c}$ where $\alpha_{c} \sim 1$ is some 
characteristic value that generally depends on $B_z$, the instability is 
relatively weak and grows on the Alfv\'en timescale characterized by the 
axial field. In contrast, if $\alpha > \alpha_{c}$, the instability is 
much more efficient, and the growth time is of the order of the Alfv\'en 
timescale for the toroidal field. 

\begin{figure}
%\begin{center}
\includegraphics[width=9.0cm]{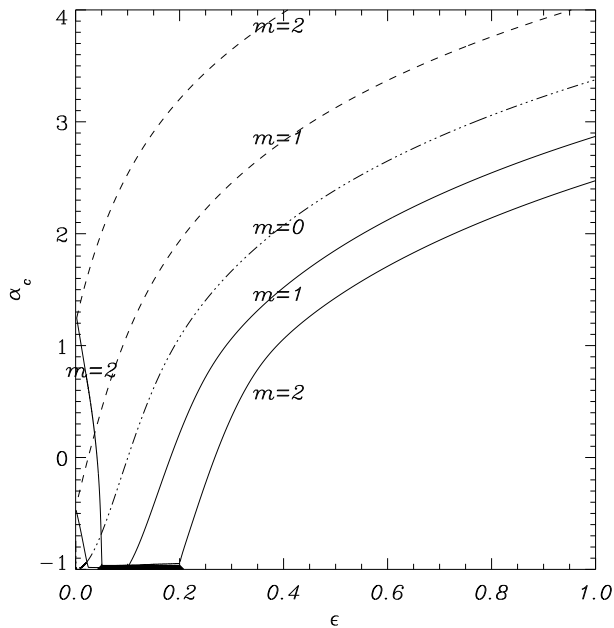}
\caption{The critical value of $\alpha$ that determines the onset of 
instability as a function of $\varepsilon$ for $|q| = 10$. Solid and dashed
curves correspond to $q=10$ and $q=-10$, respectively. The dash-and-dotted 
curve show critical $\alpha$ for $m=0$, which does not depend on the sign of 
$q$.}
%\end{center}
\end{figure}

{ In Fig.~8,  the critical value of $\alpha_{c}$ above  which  
the system is unstable is plotted as a function of $\varepsilon$ for 
%It turns out that $\alpha_{c}$ is sentitive to the allowed range of $k_z$, and 
%in particular to perturbations with very long wavelengths (small 
%$k_z$). On the other hand, as we are mainly interested in applications to stellar conditions, 
%$\alpha_c$ is calculated for 
a range of vertical wavevectors relevant to stellar conditions. 
In fact, the local approximation in the axial direction applies if 
$|k_z s| > 1$ that is equivalent $|q| > 1$. On the other hand, to neglect gravity can be justified
if $|q| > N/\omega_{B0}$ (see Eq.~(11)), which generally imposes a stronger 
restriction on $q$ since the ratio $N/\omega_{B0}$ is typically $> 1$ in 
stars. Therefore, in determining $\alpha_{c}$, we assume that the allowed $|q|$ should 
be  large enough and we choose $|q| \geq 10$, which corresponds to axial 
wavelengths shorter than $\lambda_c/10$.
%For a given $m$ and $\varepsilon$, 
%we calculated $\alpha$  that distinguish stable and unstable regimes for 
%different $|q| \geq 10$. Then, the minimal value of $\alpha$ is plotted 
%in Fig.~8 as a function of $\varepsilon$ for different $m$. 
Since the presence
of an axial field breaks symmetry between positive and negative $q$, critical 
curves are different for positive and negative wavevectors. The region of 
$\alpha$ above the lines corresponds to configurations that are unstable for 
a given $m$. It turns out that true $\alpha_{c}$, which determines instability, 
corresponds to perturbations with positive $m$ and negative $q$  as shown
in Fig.8 by solid lines. Critical $\alpha$ decreases with decreasing 
$\varepsilon$ everywhere except a region 
of small $\varepsilon$ where the dependence $\alpha_{c}(\varepsilon)$ is very 
sharp. This particularly concerns the curve $m=0$, which goes up very sharply
at $\varepsilon < 0.01$ and reaches the value 1 at $\varepsilon =0$ in 
agreement with the result by Tayler (1973). However, the scale of Fig.~8
does not let us see this sharp behaviour. Unfortunately, our code does 
not allow to follow the behaviour of critical $\alpha$ when it approaches 
the value $-1$ because of the singular character of the last term on the 
left hand side of Eq.~(15). The corresponding region is marked by crosses in Fig.~8.
A more refined consideration is needed for this case which should perhaps
include dissipation  }

\section{Discussion}
We have considered the linear stability of magnetic configurations containing 
the toroidal and axial fields, assuming that the behavior of small 
perturbations is governed by equations of the non-dissipative incompressible 
magnetohydrodynamics. This approximation is  justified if the magnetic 
field is subthermal and the Alfv\'en velocity is low compared to the sound 
speed. The stability of magnetic configurations is a key issue for 
understanding the properties of various astrophysical bodies such as peculiar 
A and B stars, magnetic white dwarfs and neutron stars. The magnetic 
instability can alter qualitatively the properties of configurations 
generated, for example, by dynamo in stars. Many dynamo models predict 
that the toroidal field should typically be stronger than the poloidal one, 
but such configurations can be unstable if the generated toroidal field does 
not decrease enough rapidly with $s$. The instability generates large- 
and small-scale motions that should alter the geometry of a generated 
magnetic field. 

Even though the poloidal field is weaker than the toroidal one in a number of 
dynamo models, its effect on the stability properties cannot be neglected. 
This  particularly concerns the behavior of the nonaxysimmetric perturbations 
considered in the present paper. If $B_z$ is relatively weak ($B_z < 
B_{\varphi}$) then, typically, there exists a wide range of the azimuthal 
wave-numbers $m$ for which the instability may occur. For any given $m$, only 
perturbations within some particular range of the vertical wave-vectors
$k_z$ can be unstable. The growth rate is maximal for perturbations with 
$k_z$ of about 
\begin{equation}
k_z \sim - \frac{m}{s} \; \frac{B_{\varphi}}{B_z}.
\end{equation}     
Equation~(19) is approximately equivalent to the condition that the Alfv\'en 
frequency is vanishing somewhere within the cylindrical layer (see Eq.~(17)). 
{It should be noted that the ratio $B_z/ B_{\varphi}$ can be rather low in stars, 
and the maximum growth rate at given $m$ corresponds to very short axial 
wavelengths,
\begin{equation}
\lambda_z \sim 2 \pi \; \frac{s}{m} \; \frac{B_z}{B_{\varphi}}.
\end{equation}
Taking, for example, $B_z/B_{\varphi} \sim 10^{-2}$ and assuming that $s$ is 
comparable to the stellar radius, $s \sim 10^{11}$ cm, we find that the 
most rapidly growing modes should have the axial wavelength $\sim 6 \times 
10^9/ m$ cm. From our results it follows that the maximum growth rate increases
slowly with increasing $m$ and, therefore, perturbations with a very short 
azimuthal wavelength (very large $m$) should dominate the development of
instability. For instance, the most rapidly growing mode with $m=100$
corresponds to the axial wavelength $\sim 6 \times 10^7$ cm, which  is very
short compared to the radius.}
Therefore, the instability of magnetic configurations  can often be determined 
by the modes with very large $m$ and an extremely short wavelength in the 
$z$-direction. This fact can cause problems in numerical modeling of 
the instability because simulations will require a very high resolution in 
the $\varphi$- and $z$-directions. 
        
Depending on the profile of the toroidal field and the strength of the 
axial field, the instability can arise in two essentially different regimes. 
In the case of a weak axial field, $B_{\varphi 0} \gg B_z$, the value of 
$\alpha$ that distinguishes between the regimes is $\approx -1/2$. If $\alpha 
> -1/2$, then the instability grows on the Alfv\'en timescale determined by 
the toroidal field and is rather fast. In this case, the growth time is
\begin{equation}
\tau \sim 0.1 \rho_{-4}^{1/2} s_{11}B_{\varphi 3}^{-1} \;\;\; {\rm yrs},
\end{equation}
where $s_{11} = s /10^{11}$cm. If $\alpha < -1/2$, then the growth 
time is given by same expression (18) but where $B_{\varphi 3}$ should be 
replaced by $B_{z 3}= B_z/ 10^3$ G. Since $B_{\varphi} \gg B_z$, the 
instability is slower and grows on the timescale determined by the axial 
field in this case. The transition between two regimes occurs at larger 
$\alpha$ if the axial field increases. 

It is fairly difficult to compare our results obtained for a simple model 
with the available numerical simulations, which usually use completely different 
basic magnetic configurations. For example, in calculations by Braithwaite 
(2006, 2007), the basic configuration was assumed to be either purely 
toroidal or purely poloidal, and the stability properties of such 
configurations differ qualitatively from those considered in this paper. 
Recently, Braithwaite \& Nordlund (2006) and Braithwaite (2008) have 
considered stability of the magnetic configuration with random initial 
fields. A vector potential was set up as a random field containing spatial 
scales up to a certain value. This random field was then multiplied by some 
screening function, so that the field strength in the atmosphere was 
negligible. 

This initial configuration contains both the toroidal and poloidal
fields but is very different from our simple model. Nevertheless, some 
features seem to be in common even for such different models. { Braithwaite
\& Nordlund (2006) and Braithwaite (2008) find that the instability can lead 
to different equilibrium configurations depending on the screening function. 
If the screening function for random fields decreases slowly or does not 
decrease at all, then the final equilibrium magnetic configuration is 
essentially non-axisymmetric. In contrast, an equilibrium configuration is
closer to axisymmetry (but not axisymmetric) if the screening function 
decreases rapidly. This dependence on equilibrium configurations obtained in
numerical calculations can reflect the regimes of ``strong'' and ``weak'' 
instabilities that correspond to different growth rates depending on the value
of $\alpha$. In accordance with our analysis, the growth rate at given 
$\varepsilon$ is higher for higher $\alpha$ (compare, e.g., Figs.~3 and 7). 
Our parameter $\alpha$ mimics to some extent the screening parameter $p$ 
introduced by Braithwaite (2008) with decreasing $\alpha$ corresponding to 
increasing $p$. Therefore, we can expect from our analysis that non-aximetric 
instabilities should be more efficient for the screening function with $p=0$ 
than with $p=1$ and that the final configuration exhibits stronger departures 
from axisymmetry for smaller $p$. This conclusion seems to be in qualitative 
agreement with the results of Braithwaite (2008).}

Our simple model does not take into account the stratification that can be
important in many astrophysical applications. Basically, stratification 
provides a stabilizing influence if the temperature gradient is sub-adiabatic.
However, this influence is small if perturbations have a relatively short
wavelength in the axial direction, $\lambda < \lambda_c$, such that inequality 
(11) is satisfied. Our results are related to this case. The case when 
$\lambda > \lambda_c$ and stratification is important will be considered 
elsewhere.

\vspace{0.5cm}

\noindent

{\it Acknowledgments.}
This research project was  supported by a Marie Curie Transfer of
Knowledge Fellowship of the European Community's Sixth Framework
Program under contract number MTKD-CT-002995.
VU also thanks  INAF-Ossevatorio Astrofisico di Catania for hospitality.

{}

\end{document}